\documentclass[a4paper]{ar-1col}
\usepackage{graphicx}
\usepackage[numbers]{natbib}

\usepackage{graphicx}
\usepackage{amssymb}
\usepackage{amsmath}
\usepackage{amsfonts}
\usepackage{color}
\usepackage{dsfont}
\usepackage{paralist}
\usepackage{bm}
\usepackage{verbatim}
\usepackage{textcomp}
\usepackage{float}

\jname{Ann. Rev. Cond. Mat. Phys.}
\jvol{9}
\jyear{2017}
\doi{10.1146/((article doi))}

\def\cH{\hat{\cal H}}

\def\cL{{\cal L}}
\def\cO{{\cal O}}

\def\bk{{\bf k}}

\def\br{{\bf r}}

\def\hbsigma{\hat{\boldsymbol \sigma}}

\DeclareMathOperator{\sign}{sign}

\begin{document}

\markboth{S.V.~Syzranov and L.~Radzihovsky}{Non-Anderson Transitions}

\title{High-Dimensional Disorder-Driven Phenomena in Weyl Semimetals, Semiconductors and Related Systems}

\author{Sergey V. Syzranov$^{1,2}$ and Leo Radzihovsky$^{1,3,4}$
\affil{$^1$Physics Department and Center for Theory of Quantum Matter,\\
University of Colorado, Boulder, USA, CO 80309; \\
email: sergey.syzranov@googlemail.com}
\affil{$^2$Joint Quantum Institute, NIST/University of Maryland,\\
College Park, USA, MD 20742 
}
\affil{$^3$JILA, University of Colorado, Boulder, USA, CO 80309}
\affil{$^4$KITP, University of California, Santa-Barbara, USA, CA  93106}}

\begin{abstract}
It is commonly believed that a non-interacting disordered electronic system 
can undergo only the Anderson metal-insulator transition.
It has been suggested, however, that a broad class of systems can display disorder-driven
transitions distinct from Anderson localisation that have manifestations in the disorder-averaged density
of states, conductivity and other observables. Such transitions have received particular attention
in the context of recently discovered 3D Weyl and Dirac materials but have also been predicted in cold-atom
systems with long-range interactions, quantum kicked rotors, and all sufficiently high-dimensional systems.
Moreover, such systems exhibit unconventional behaviour of Lifshitz tails, energy-level statistics,
and ballistic-transport properties. Here we review recent progress and the status of results
on non-Anderson disorder-driven transitions and related phenomena. 
\end{abstract}

\begin{keywords}
localisation, phase transitions, disordered systems, Dirac semimetals
\end{keywords}
\maketitle

\tableofcontents

\section{INTRODUCTION}

Quenched disorder, such as impurities, vacancies, and dislocations, is inherently present in all
solid-state systems. It determines conduction, affects the properties of phases and their associated phase transitions,
and leads to new phases of matter.

The amount of quenched disorder often determines the difference between metals and insulators, respectively
materials with finite and zero conductivity at zero temperature. When the amount of disorder is increased,
a metal can turn into an insulator due to the Anderson localisation\cite{Anderson:original}
of electron states at the Fermi surface. This insulating behaviour at strong disorder
persists, in particular, in the presence of sufficiently weak interactions\cite{BAA}. 

Interest in localisation phenomena and the other effects of disorder
continues to be fuelled by discoveries of new materials with non-trivial
topologies and quasiparticle band structures, including $d$-wave superconductors, graphene, 
and 3D Dirac and Weyl semimetals. 

The novel band structures and topological properties in
such materials require a reexamination of transport and thermodynamic phenomena.
While in systems with large electron density (with the Fermi level deep inside the band), one can apply the conventional treatment based on an effective constant density of states, elastic scattering rate, electron mass etc.\cite{Abrikosov:metals,AGD,Efetov:book}, such an approach 
fails completely when the density of states vanishes, which occurs near nodes
(e.g., in graphene
or Weyl semimetals) and band edges. 
In these regimes, which are the subject of this review, a qualitatively different treatment is required and new phenomena are expected and indeed found. 

For example, under certain approximations it was demonstrated\cite{Fradkin1,Fradkin2} more than 30 years ago that a three-dimensional (3D) system with Dirac quasiparticle dispersion exhibits a novel disorder-driven transition
between two phases, with zero and finite densities of states (DoS), at the Dirac node. This contrasts qualitatively with
the Anderson localisation transition, commonly believed to be the only possibility for a non-interacting disordered system, where the DoS exhibits no interesting features through the transition.

Interest in the prediction of this unconventional (non-Anderson) disorder-driven phase
transition has been revived recently by the advent of 3D
Dirac\cite{Young:BiO2Diracproposal,Wang:A3BiDiracproposal,Liu:Na3BiDiracdiscovery,Liu:Cd3As2Diracdiscovery} and
Weyl\cite{Wan:WeylProp,ZHasan:TaAs,ZHasan:TaAs2,ZHasan:TaP,ZHasan:NbAs,Weng:PhotCrystWSM} semimetals (WSMs).
Extensive theoretical and numerical studies [see, e.g., Refs.~\cite{ShindouMurakami,RyuNomura,Goswami:TIRG,Herbut,OminatoKoshino,Syzranov:Weyl,Syzranov:unconv,
Brouwer:WSMcond,Pixley:twotrans,Pixley:ExactZpubliahed,Chen:WSM-QAH,LiuOhtsuki:LateNumerics,Brouwer:exponents,
Shapourian:PhaseDiagr,Bera:Weyl,Syzranov:twoloopZ,PixleyHuse:rare,RoyDasSarma:erratum,
Roy:ac,Syzranov:multifract,Carpentier:phi4extraboson,PixleyHuse:rare2}] over the last several years have
been directed at exploring the properties of such transitions and the behaviour of physical
observables in Dirac and Weyl systems. Although the density of states never exactly
vanishes\cite{Wegner:DoS,Suslov:rare,Nandkishore:rare,Skinner:WeylImp,Syzranov:unconv},
in contrast with the expectations of the pioneering studies\cite{Fradkin1,Fradkin2},
this nevertheless does not preclude a singular (non-analytic) behaviour near a critical
value of the disorder strength on top of a smooth non-zero background.

Alternatively, it has been argued [see, e.g., Refs.~\cite{PixleyHuse:rare} and \cite{PixleyHuse:rare2}] that the proposed non-Anderson 
transition may in fact be a sharp crossover, which is rounded off in a small region in the vicinity of
the putative critical point by rare-region effects, as we will discuss in Section~\ref{Sec:rare}.
In most of this review, however, we will not distinguish between such a sharp crossover and a true disorder-driven
phase transition and refer to the proposed critical point as ``non-Anderson disorder-driven transition''.
\begin{figure}[htbp]
	\centering
	\includegraphics[width=0.7\textwidth]{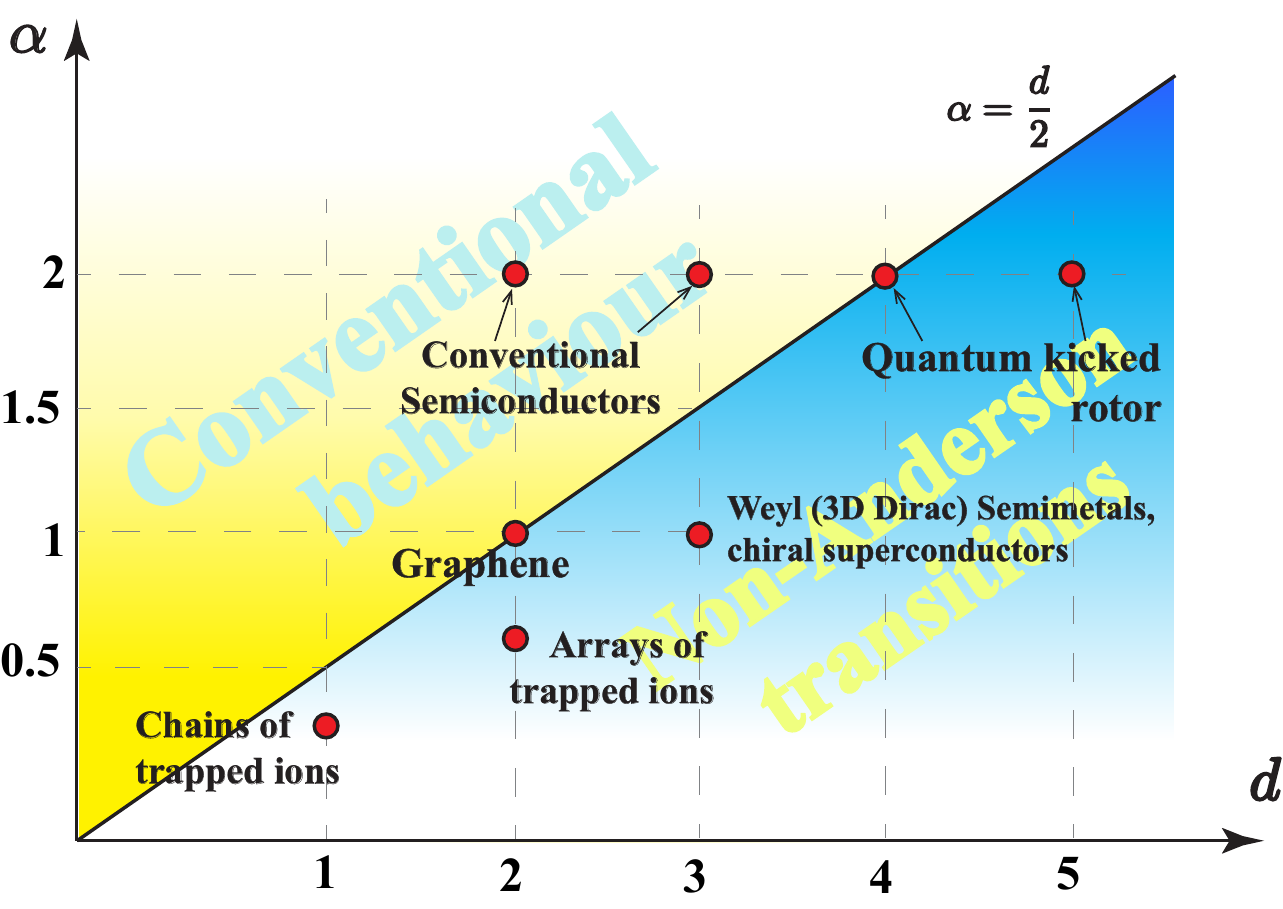}
	\caption{\label{PeriodicTable}
	Examples of systems with the power-law quasiparticle dispersion $\epsilon_\bk\propto k^\alpha$.
	Systems in high dimensions, $d>2\alpha$, exhibit non-Anderson disorder-driven transitions.
	}
\end{figure}
{Although the DoS is not an order parameter for this transition 
in a strict sense, it can be used to distinguish the phases under certain approximations; 
the DoS is suppressed for disorder strengths weaker than a critical value and grows
rapidly for stronger disorder.}
{In the case this non-Anderson transition is a true phase transition, which is currently
not established, the phases may be distinguished by a qualitative change in the behaviour
of correlation functions of physical observables.}

It has been demonstrated recently\cite{Syzranov:Weyl,Syzranov:unconv} that
such a transition in 3D Dirac and Weyl materials
is a special case of a more general non-Anderson disorder-driven transition that occurs in all systems
in sufficiently high dimensions, summarised by a "phase diagram" with a number of examples in Fig.~\ref{PeriodicTable}. 
In particular, such transitions take place in all systems with the power-law quasiparticle
dispersion $\epsilon_\bk\propto k^\alpha$ in dimensions $d>2\alpha$ (referred to in this review 
as ``high dimensions''), as illustrated in Fig.~\ref{PeriodicTable}.
As we discuss in this review, the best known example here, of direct experimental relevance,
is 3D Dirac materials ($\alpha=1$, $d=3$), such as Dirac and Weyl semimetals and chiral superconductors.
The transition has also been predicted to be realisable\cite{Garttner:longrange} in
1D\cite{Monroe:longrange,Islam:longrange,Blatt:chain1,Blatt:chain2}
and 2D\cite{Bollinger:longrange} arrays of
trapped ions with long-range interactions; such arrays host quasiparticles with the power-law dispersion
$\varepsilon_\bk\propto k^\alpha$ with tunable $\alpha$ and thus present a flexible platform for studying the transition.
Another class of systems, which allow for the observation of non-Anderson
transitions, is quantum rotors kicked by perturbations 
with $d$ incommensurate frequencies\cite{Moor:rotorrealisation,Chabe:rotorrealisation,Delande:rotorrealisation}:
such a rotor can be mapped\cite{Grempel:mapping,Casati:mapping}
onto a disordered semiconductor in dimension $d$ {and is thus expected to display a transition for $d>4$.}
{Quantum kicked rotors with arbitrary $d$ may be realised in cold-atom systems and have already been demonstrated
for $d=1$ \cite{Moor:rotorrealisation} and $d=3$ \cite{Chabe:rotorrealisation,Delande:rotorrealisation}.}
High-dimensional disorder-driven transitions
can be also studied numerically in arbitrary
dimensions\cite{Markos:review,GarciaGarcia,Slevin,Zharekeshev:4D,SlevinOhtsuki:HighDUnitary}.

\begin{figure}[htbp]
	\centering
	\includegraphics[width=0.95\textwidth]{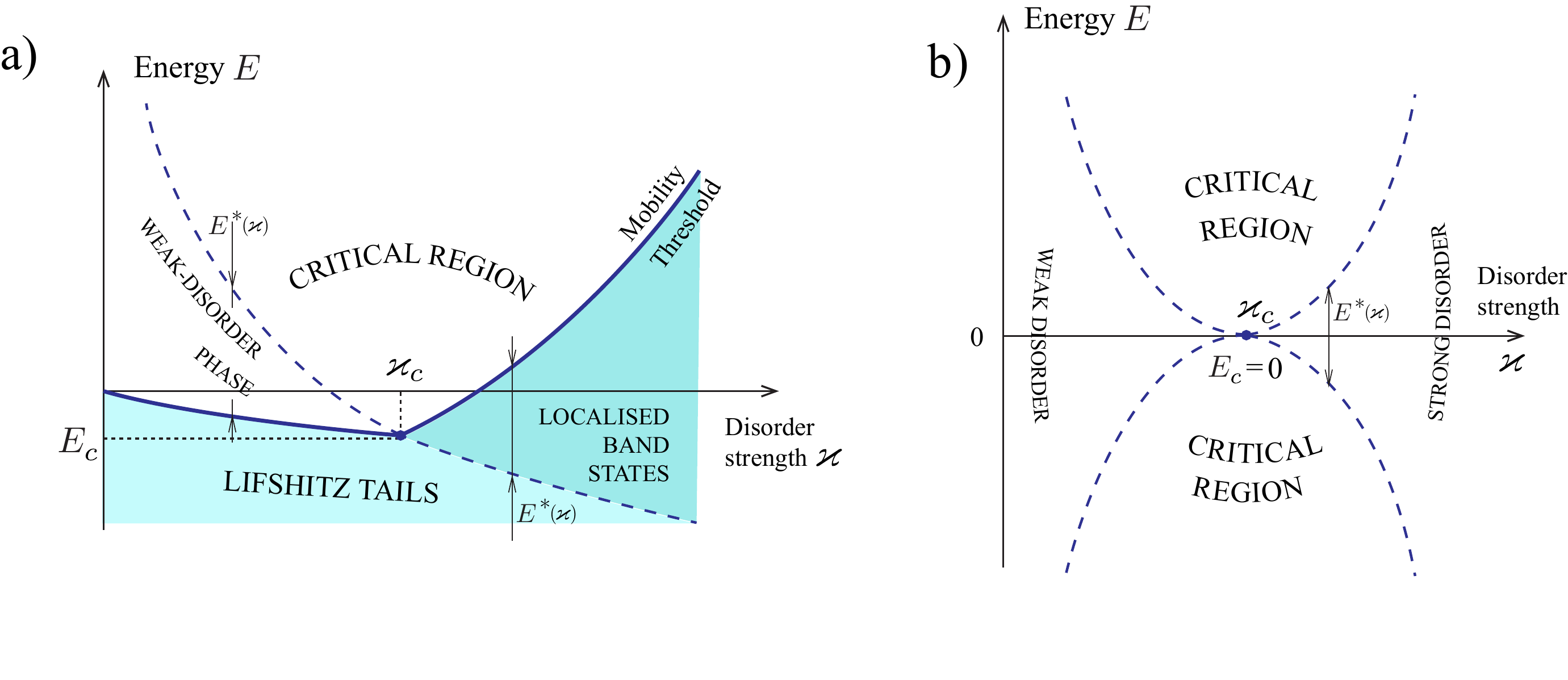}
	\caption{\label{PhaseDiagrSemicond}
	Phase diagram of materials that exhibit non-Anderson disorder-driven transitions:
	a) ``high-dimensional'' semiconductor in the orthogonal symmetry class, b) Weyl semimetal and 3D Dirac materials.
	The transition takes place at a single energy, $E_c$ (in a WSM $E_c=0$). 
	In a semiconductor the transition is expected to lead to non-analytic behaviour of the mobility
	threshold; it is pinned to the band edge (neglecting Lifshitz tails) for subcritical disorder
	and grows rapidly for supercritical disorder.	
	}
\end{figure}

This disorder-driven transition occurs for single-particle states only at one value of energy, $E_c$,
near a node or a band edge, Fig.~\ref{PhaseDiagrSemicond}, in contrast to
the Anderson localisation transition, which occurs
for any energy in the band when the disorder strength is varied.
This non-Anderson transition
can also be accompanied by a localisation-delocalisation of the wavefunctions
at the transition point, if allowed by symmetry, dimension, and topology.
However, these transitions also occur in systems
where localisation is forbidden by symmetry and/or topology, such as 1D chiral chains\cite{Garttner:longrange}
and single-node Weyl semimetals,
and in systems with all states localised (e.g., systems in the orthogonal symmetry class in dimensions
$d<2$). In this case the transition manifests itself in the critical behaviour of the DoS and other observables,
as discussed in detail below.

The purpose of this article is to review the properties of the unconventional non-Anderson
disorder-driven transitions, the status of various results associated with them, and the possibility
of their experimental observation in realistic systems.

\section{DISORDER-INDUCED PHENOMENA IN HIGH DIMENSIONS}

In this review we focus on the properties of $d$-dimensional systems described by the (single-particle) 
Hamiltonian
\begin{align}
	\cH=a|\bk|^\alpha + U(\br),
\end{align}
where $\bk=-i\partial_\br$ is the momentum operator {(throughout this review
we set $\hbar=1$)}, $U(\br)$ is a short-range-correlated random
potential with zero average,
and $\epsilon_\bk=a|\bk|^\alpha$ is the kinetic energy.
In some materials the kinetic energy has an additional spin or valley structure;
for instance, in graphene and Weyl semimetals
$\epsilon_\bk=v\bk\cdot\hbsigma$, where $\hbsigma$ is a vector of Pauli matrices
in the space of a discrete degree of freedom (e.g. spin or valley space).
In most of the review 
we consider effects on length scales significantly longer than the correlation length of the random
potential, which can thus be taken to have the delta-correlated form,
\begin{align}
\langle
U(\br)U(\br^\prime)\rangle_{\text{dis}}=\varkappa\delta(\br-\br^\prime),
\label{DisorderShortStrength}
\end{align}
where the delta-function is regularised by a cutoff set by the correlation length of the random potential.

\subsection{Phenomenological argument for a 
transition near nodes and band edges}
\label{Subsec:phenom}

Before delving into technical details, we present a physical argument that suggests the
existence of the non-Anderson disorder-driven  phase transition in sufficiently high dimensions.

The effective strength of a weak random potential with a typical 
amplitude $U_0\ll ar_0^{-\alpha}$ and correlation length $r_0\ll 1/k$ can be 
characterised by the average of the potential over the de Broglie wavelength $2\pi/k$. 
According to 
the central limit theorem, this gives $U_{\text{rms}}\sim U_0(kr_0)^\frac{d}{2}$, which for low momentum states $k$
in high dimensions $d > 2\alpha$ is always subdominant to the corresponding kinetic energy $a k^\alpha$. 
By contrast, in low dimensions $d < 2\alpha$ an arbitrarily weak random potential always dominates at sufficiently long wavelengths, which leads, for example, to the localisation of states at the bottom of the band
of a conventional semiconductor.

This strict {\em perturbative} argument breaks down for strong random potential, which
is expected to modify the above weak-disorder state qualitatively and make the system behave
similarly to a conventional semiconductor.
Taken together, this thus suggests a disorder-driven transition between weak- and strong-disorder phases. 

We emphasise that these arguments apply to all systems in high dimensions $d>2\alpha$ and are
not specific to Dirac materials where the transition was first suggested\cite{Fradkin1,Fradkin2}.
{Moreover, this transition is a priori unrelated to the Anderson localisation transition or even to its
existence in the system under consideration. Indeed, the above arguments 
apply regardless of whether disorder allows for coherent backscattering required for Anderson localisation.
Also, the physics of this high-dimensional transition is controlled by
length scales shorter than the mean free path $\ell$,
while the properties of Anderson localisation and its existence are determined by length scales longer than
$\ell$ \cite{Efetov:book}.}

\subsection{Self-consistent Born approximation ($N=\infty$ approach)}

We now turn to a more rigorous quantitative description of the non-Anderson transition and
physical observables near it.

{\it Born approximation.} The above physical arguments are reflected in the behaviour of the
elastic scattering rate $\Gamma(E)$ as a function of the particle energy $E$. In the leading-order perturbation theory (Born approximation) $\Gamma(E)\propto E^{\frac{d}{\alpha}-1}$. 
The dimensionless strength of disorder $\gamma(E) = \Gamma(E)/E \equiv [k\ell(k)]^{-1}
\propto E^{\frac{d}{\alpha}-2}$
shows that in high dimensions $d > 2\alpha$ effects of weak disorder only become weaker at the bottom of 
the band, $E\rightarrow 0$. 

By contrast, for $d<2\alpha$ disorder cannot be treated perturbatively at low energies,
which reflects, for instance, in the localisation of states near semiconductor band edges.

\vspace{2mm}
\noindent
{\it Self-consistent Born approximation.}
To treat strong disorder one can try to employ the self-consistent Born approximation (SCBA),
that was used to first predict\cite{Fradkin1,Fradkin2} the transition in Dirac
materials 30 years ago. 

The approximation  amounts to a self-consistent evaluation of the scattering rate
$\Gamma$ using the Green's function, e.g., $G^R=(E-\varepsilon_\bk+i\Gamma/2)^{-1}$, that contains the same scattering rate as a function of the disorder strength $\varkappa$ and energy $E$:
\begin{align}
	\Gamma(\varkappa,E)=\varkappa\int\frac{d^d\bk}{(2\pi)^d}\frac{\Gamma(\varkappa,E)}
	{(E-ak^\alpha)^2+[\Gamma(\varkappa,E)]^2/4}.
	\label{SCBAeq}
\end{align}
Eq.~(\ref{SCBAeq})
applies both to semiconductors with the quasiparticle dispersion $\epsilon_\bk=a k^\alpha$ and to
systems with dispersions that have more complicated spin or valley structures, e.g. Weyl semimetals,
such that $|\epsilon_\bk|=ak^\alpha$. Above for simplicity we disregard the real part of the self-energy
responsible for the disorder-induced shift of the lowest energy levels.

The self-consistency equation (\ref{SCBAeq}) matches the saddle-point
solution of the non-linear sigma-model\cite{Efetov:book}.
Fluctuations around the saddle-point solutions make physical observables deviate from the SCBA result.
Such fluctuations, however, can be taken into account systematically via the $1/N$ expansion
in a generalised model with $N$ valleys (particle flavours) between which disorder can scatter.
SCBA results thus become exact in the limit $N\rightarrow\infty$.

For states with $E=0$, Eq.~(2) has only the trivial $\Gamma=0$ solution for disorder strength weaker than the critical value
\begin{align}
	\varkappa_c^{SCBA}=\left[\int_{|\bk|<K_0}\frac{d^d\bk}{(2\pi)^d}\frac{1}	{a^2k^{2\alpha}}\right]^{-1},
	\label{SCBAcritDisorder}
\end{align}
while for $\varkappa > \varkappa_c$ there is a nontrivial $\Gamma\neq 0$ solution,
where $K_0$ is the ultraviolet momentum cutoff set by the bandwidth or the inverse impurity size $r_0^{-1}$.
Thus, the
SCBA predicts a disorder-driven phase transition at the critical disorder strength $\varkappa_c$,
Eq.~(\ref{SCBAcritDisorder}) (provided it is non-zero) between phases with $\Gamma = 0$ and $\Gamma\neq 0$.

In the case of low dimensions, $d<2\alpha$, $\varkappa_c^{SCBA}=0$, suggesting a finite scattering rate
for all disorder strengths and the absence of the phase transition.
For high dimensions ($d>2\alpha$), the critical disorder strength
$\varkappa_c^{SCBA}$ is non-zero, signalling a possible phase transition.

The SCBA and the (equivalent in the leading order) large-$N$
approach have also been used to analyse disorder-driven criticality for 3D Dirac fermions in 
Refs.~\cite{ShindouMurakami} and \cite{RyuNomura}. In Ref.~\cite{OminatoKoshino} the SCBA was
used to compute the conductivity near the non-Anderson disorder-driven transition in
a Weyl semimetal.

\subsubsection*{Criticism of the SCBA.}
While the SCBA is justified in the limit of an infinite number of valleys, $N\rightarrow\infty$,
it becomes inaccurate for realistic materials with finite $N$.
The SCBA has been criticised in the context of graphene in Ref.~\cite{AleinerEfetov}
and also later discussed in detail in Ref.~\cite{OstrovskyGornyMirlin}. 
Although graphene corresponds to the marginal dimension $d=2\alpha$ and does not exhibit a transition,
the criticism of Refs.~\cite{AleinerEfetov} and \cite{OstrovskyGornyMirlin} also applies to all
systems in higher dimensions $d\geq2\alpha$ [a detailed analysis of the SCBA for higher-dimensional systems
and Weyl semimetals may be found in Refs.~\cite{Syzranov:Weyl} and \cite{Brouwer:WSMcond} respectively].

When applied to realistic systems, the SCBA
takes into account certain contributions to physical observables, e.g. the density of states and
conductivity, and systematically neglects other contributions of the same order of magnitude\cite{AleinerEfetov,OstrovskyGornyMirlin,Syzranov:Weyl,Brouwer:WSMcond}.
Thus, controlled methods, to which we turn in the next section,
are necessary to analyse more systematically the nature of the
disorder-driven transition.

\subsection{Renormalisation-group analysis}

\label{Sec:RG}

\begin{figure}[htbp]
	\centering
	\includegraphics[width=0.6\textwidth]{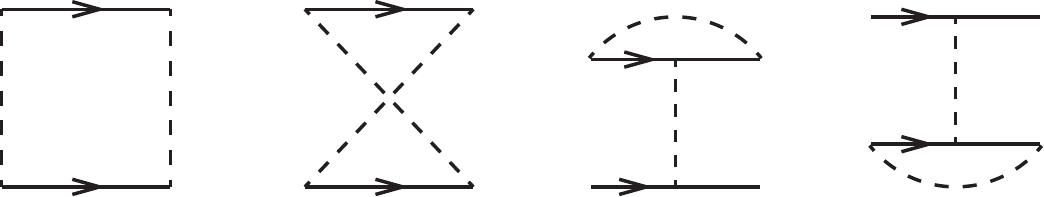}
	\caption{\label{Diagrams}
	Diagrams for the one-loop renormalisation of the disordered strength. 
	}
\end{figure}

Non-Anderson disorder-driven transitions 
can
also be analysed using a more systematic 
renormalisation group (RG) approach\cite{GellManLow:firstRG,Wilson:firstRG,Sachdev:book}, which is controlled
by the small parameter $\varepsilon=2\alpha-d$ \cite{Syzranov:Weyl}.
The celebrated approach consists in coarse-graining the system by integrating out the high-momentum
degrees of freedom down to an ultraviolet momentum cutoff $K=K_0 e^{-l}$, thereby
renormalising the system properties at lower momenta,
where $K_0$ is set by the bandwidth or disorder correlation length.
For simple potential disorder, which is the most common in realistic systems,
the associated RG flow of 
the dimensionless disorder strength, 
$\gamma(l)\propto \varkappa K^{-\varepsilon}$, is given by
\begin{align}
	\partial_l\gamma=(2\alpha-d)\gamma+\gamma^2+\ldots.
	\label{RGeq}
\end{align}
 We note that the dimensionless disorder strength $\gamma$ is of the order of 
the inverse Ioffe-Regel parameter\cite{Syzranov:unconv} $(k\ell)^{-1}$, which plays an important role in the theory
of transport and localisation in disordered systems\cite{Abrikosov:metals,AGD,Efetov:book}.
The first term in the right-hand side of Eq.~(\ref{RGeq}) arises from the power-counting for
the random perturbation at the Gaussian disorder-free fixed point. The second term, quadratic-in-disorder,
arises from the one-loop corrections to the strength of disorder corresponding to the diagrams in Fig~\ref{Diagrams}.

In accordance with the arguments of Sec.~\ref{Subsec:phenom},
Eq.~(\ref{RGeq}) shows that weak disorder strength always grows at low momenta in low dimensions ($d<2\alpha$).
However, slightly above the marginal dimension, $d_c=2\alpha$, i.e.
for small negative values of $\varepsilon=2\alpha-d$, the disorder strength flows to larger or smaller values,
depending on whether or not it exceeds the critical value $\gamma_c=|\varepsilon|$.
This signals the aforementioned disorder-driven quantum phase transition.

Some realistic systems, such as arrays of (optically or magnetically) trapped
ions\cite{Monroe:longrange,Islam:longrange,Bollinger:longrange,Blatt:chain1,Blatt:chain2} can be tuned\cite{Garttner:longrange} to realise arbitrarily small values of $\varepsilon$.
Other systems, such as semiconductors or Weyl semimetals, can realise only integer values of $\alpha$
and $d$, leading to integer values of $\varepsilon$ of order unity.

Although the above RG analysis is controlled quantitatively only for small $\varepsilon$, 
we expect that the phase transition and its qualitative properties persist to values of $\varepsilon$ of order $1$,
assuming the absence of a phase transition as a function of $\varepsilon$ (which, strictly speaking,
cannot be excluded). Indeed, the numerical results for the DoS in
Weyl (Dirac) semimetals\cite{Herbut,LiuOhtsuki:LateNumerics,Brouwer:exponents,Pixley:ExactZpubliahed,Bera:Weyl}
and in chains of trapped ions\cite{Garttner:longrange} agree well with 
the one-loop RG analysis for $|\varepsilon|\sim1$, {as we will discuss in more detail in Sec.~\ref{Sec:observables}}.

An RG approach of this type was developed nearly three decades ago\cite{DotsenkoDotsenko} to
describe 2D Dirac fermions
in the context of the Ising model. Later it was applied to 2D Dirac quasiparticles
in integer-quantum-Hall systems\cite{LudwigFisher} and $d$-wave superconductors\cite{Bocquet:2DRG,AltlandSimonsZirnbauer}
[see also Ref.~\cite{Guruswamy:2DRG}]. 
More recently, a comprehensive RG description for electrons in graphene with various disorder
symmetries has been developed in Ref.~\cite{AleinerEfetov} and further discussed in Ref.~\cite{OstrovskyGornyMirlin}.
Although 2D Dirac quasiparticles experience renormalisation effects from elastic scattering
through all momentum states corresponding to the linear dispersion, they correspond to the critical
dimension $d=2\alpha=2$ ($\varepsilon=0$) and do not exhibit the transition.

A non-Anderson disorder-driven transition was first studied in this type of RG framework for
3D fermions ($\alpha=1$, $d=3$) in Ref.~\cite{Goswami:TIRG}, also taking into account the electron-electron interaction.
Such approaches are now being
used very broadly to study the properties of the transition and critical behaviour of physical observables
in Weyl semimetals [see, e.g., 
Refs.~\cite{Syzranov:Weyl,Syzranov:unconv,Pixley:twotrans,Pixley:ExactZpubliahed,Bera:Weyl,Syzranov:twoloopZ,RoyDasSarma:erratum,
Roy:ac,Syzranov:multifract,Carpentier:phi4extraboson}], where again $\alpha=1$ and $d=3$.
In Refs.~\cite{Syzranov:Weyl} and \cite{Syzranov:unconv} it was demonstrated that such non-Anderson transitions
are not specific to Dirac and Weyl systems, but exist in all systems in high dimensions, $d>2\alpha$,
as can be seen when the RG analysis is applied to generic semiconductor-like  
and high-dimensional chiral systems.
In particular, in Ref.~\cite{Garttner:longrange} this RG analysis was applied
to 1D spin chains mapped onto the problem of a single particle with a power-law dispersion
in a random potential, leading to the prediction of the unconventional disorder-driven phase transition
in such systems.

\subsection{Non-Anderson vs. Anderson transitions}

We emphasise that the disorder-driven transition discussed in the previous subsections
is distinct from the conventional Anderson metal-insulator transition. While the former manifests itself
in physical observables only near nodes and band edges, the usual Anderson localisation
occurs for all states away nodes and band edges (and is usually observable in the conductivity, which is
determined by the states at the Fermi level).

\begin{figure}[htbp]
	\centering
	\includegraphics[width=0.8\textwidth]{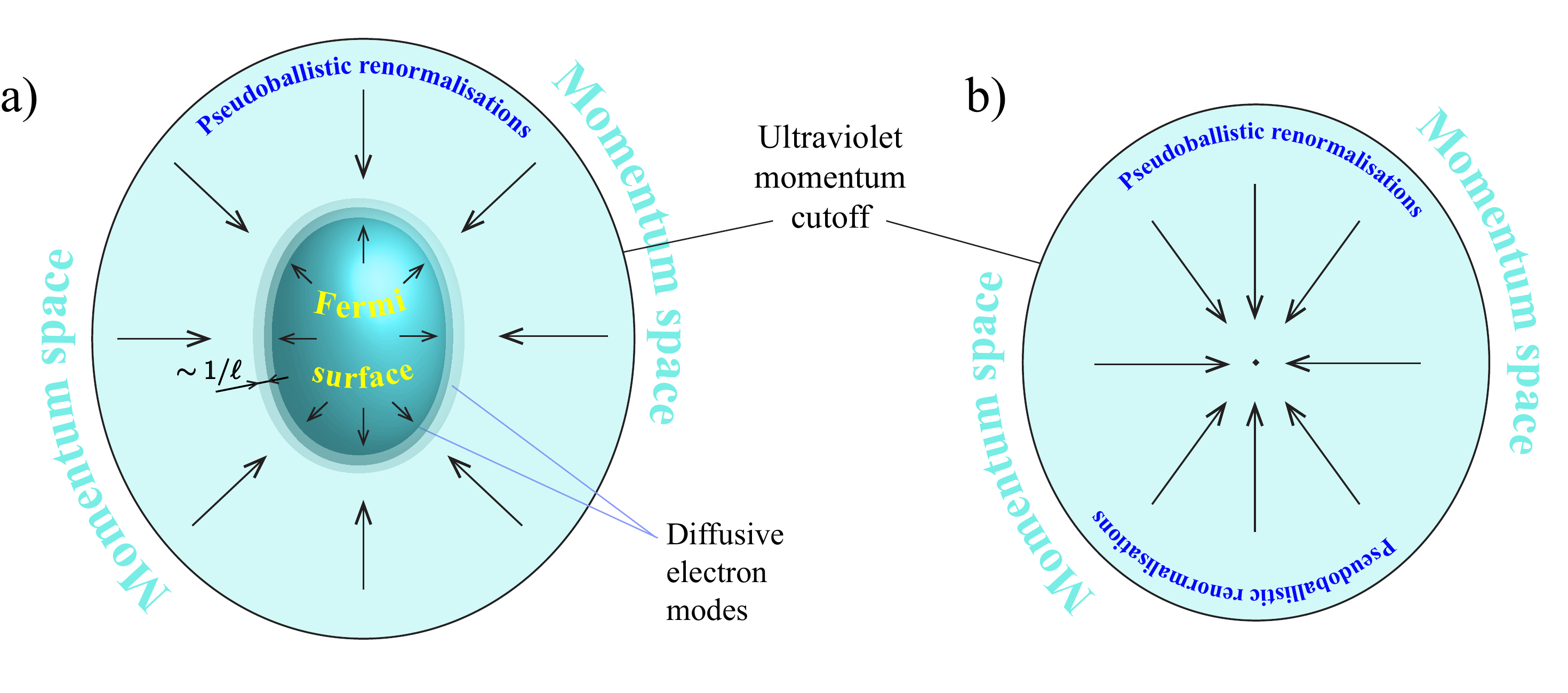}
	\caption{\label{SurfaceRG} 
	Pseudoballistic (on scales shorter than the mean free path $\ell$) renormalisation in
	high-dimensional ($d>2\alpha$) systems.
	a) For electrons at the Fermi energy, the RG is cut off by the Fermi surface. 
	The resulting renormalised parameters
	are an input for describing diffusive electron modes in the momentum shell of the 
	width $\sim1/\ell$ near the Fermi surface.
	b) For electrons near a band edge or a node, the pseudoballistic RG flow persists to the lowest momenta.
	The effective ultraviolet momentum cutoff is determined by the correlation length of the
	random potential or the band width in momentum space.	
 	}
\end{figure}

In the presence of a finite Fermi surface, one may introduce, in the spirit of the
Fermi-liquid theory\cite{Abrikosov:metals,AGD,Efetov:book}, effective (measurable in experiment)
constant DoS, electron mass and elastic scattering time near the Fermi surface. 
It is usually assumed that
scattering processes through momenta far from the Fermi surface (see Fig.~\ref{SurfaceRG})
can be absorbed in these effective parameters of the low-energy quasiparticles.
The single-particle physics is then described by the non-linear sigma-model\cite{Efetov:book}
in terms of diffusive electron modes on distances larger than the mean free path $\ell$.
This model is believed to exhibit only one disorder-driven transition, namely the Anderson localisation transition.
The critical properties of
the Anderson transition are believed to be universal, depending only on
the dimensionality of space, underlying symmetry class, and topology, but independent of the
microscopic details. In particular, in dimensions $d=2+\varepsilon$, the Anderson criticality can
be described analytically\cite{EfetovLarkinKhmelniskii,Efetov:book} using an RG approach, controlled by the
small parameter 
$\varepsilon>0$, in the framework of the
non-linear sigma-model.
The renormalisations considered in Sec.~\ref{Sec:RG} determine system properties on length scales shorter
than the mean free path $\ell$ \cite{Syzranov:unconv} and may be termed ``pseudoballistic''. 
These renormalisations are cut off by the Fermi momentum, so long as $k_F\ell(k_F)\gtrsim1$, and thus
determine only the input parameters of the sigma-model, but do not affect the Anderson criticality.

However, as we have seen above, the phenomenology is very different
near nodes and band edges, where the quasiparticle parameters
(density of states, scattering rate, etc.) can vanish or
demonstrate non-analytic behaviour due to the renormalisation by high-momentum modes,
leading to the failure of the conventional sigma-model description.
As a consequence, one may expect non-Anderson
disorder-driven transitions near nodes and band edges.
{As we discussed in the previous subsections, such unconventional transitions
are indeed found in systems in high dimensions.}

\subsection{Related phenomena in disorder-free interacting systems}

The effective action, describing a single particle with energy $E$ 
and dispersion $\epsilon_\bk=ak^\alpha$ in a random short-range-correlated
potential, in the supersymmetric\cite{Efetov:book} and replica\cite{BelitzKirkpatrick} representations is given by
\begin{align}
	\cL=-i\int_\br \bar\psi\left(E-\varepsilon_\bk+i0\cdot\Lambda\right)\psi
	+\frac{1}{2}\varkappa\int_\br(\bar\psi\psi)^2,
	\label{Action}
\end{align}
with $\bar\psi$ and $\psi$ being supervectors\cite{Efetov:book} or
vectors with components in replica space\cite{BelitzKirkpatrick}, respectively.
A similar action can also be written in Keldysh representation\cite{Kamenev:book}. 
As the action (\ref{Action}) resembles that of the $\phi^4$ theory, 
the critical behaviour at the non-Anderson disorder-driven transition
resembles that of a number of disorder-free interacting systems.

A prominent example here is the field theory of a self-avoiding polymer in $d$ dimensions\cite{deGennes:book}
with interaction strength $g$.
The RG flow of the coupling $g$ is also given by
Eq.~(\ref{RGeq}) with $\gamma\rightarrow -g$ and $\alpha=2$. 
The fixed point characterising the anomalous long-scale properties of a self-avoiding polymer appears
in $d<4$ dimensions.

Phase transitions 
driven by high momenta or energies 
are known to occur also 
in dilute weakly interacting
\cite{Singh:BoseGas1,Singh:BoseGas2,Uzunov}
and Feshbach-resonant Bose gases\cite{Radzihovsky:Feshbach,RomansSachdev:Feshbach,Radzihovsky:FeshbachReview},
as well as in
fermionic gases
near unitarity\cite{NikolicSachdev,VeiletteRadzihovsky,GurarieRadzihovsky:bcsbec} (BCS-BEC crossover). 
Similar renormalisations of low-energy quasiparticle properties due to scattering
through high-energy states occur in strongly disordered bosonic insulators\cite{Syzranov:MFRG,Syzranov:accond}, with a similar critical
point for the DoS increasing at low energies.

\section{PHYSICAL OBSERVABLES NEAR THE TRANSITION}
\label{Sec:observables}

\subsection{Critical exponents}

Under the assumption of single-parameter scaling, a continuous phase transition is characterised
by a single divergent correlation length, $\xi\propto |\varkappa-\varkappa_c|^{-\nu}$,
at the critical energy $E=E_c$, with a universal
correlation-length exponent $\nu$, 
and the {characteristic energy
$|E-E_c|=E^*\propto\xi^{-z}$}, where $z$ is the universal dynamical exponent. 
The scale $E^*(\varkappa)\propto(\varkappa-\varkappa_c)^{z\nu}$ serves
as a crossover energy between several distinct types of behaviour
of conductivity, DoS, and other physical observables, as illustrated in Fig.~\ref{PhaseDiagrSemicond}.
When Anderson localisation is allowed by symmetry in a high-dimensional system, we expect the scale $E^*$
to determine the position of the mobility threshold for $\varkappa>\varkappa_c$.
{The critical exponents can be observed experimentally, for example,
through the critical behaviour of the DoS and conductivity, as we discuss in Secs.
~\ref{Sec:DoS} and \ref{Sec:cond} below.}

The correlation-length exponent $\nu=1/(d-2)$ for $d$-dimensional Weyl fermions has been
computed in the SCBA ($N=\infty$ approach) in Ref.~\cite{Fradkin1} and has also been evaluated at
next-to-leading order in $1/N$, together with the dynamical exponent $z$, in Ref.~\cite{RyuNomura}.
For 3D Dirac and Weyl semimetals, the one-loop RG analysis predicts $\nu=1$, $z=3/2$,
result first obtained in Ref.~\cite{Goswami:TIRG}. 
Near the band edge of a semiconductor the one-loop critical exponents are given by
$\nu=(d-2\alpha)^{-1}$ and $z=\alpha/2+d/4$
for arbitrary $\alpha$ and $d$ \cite{Syzranov:unconv}.

Field theories of Weyl particles in a random potential [in replica\cite{BelitzKirkpatrick}, supersymmetric\cite{Efetov:book}, or Keldysh\cite{Kamenev:book} formulations] can be
mapped onto the Gross-Neveu model\cite{GrossNeveu} in the limit of 
a vanishing number of fermion flavours, $N\rightarrow0$, and with the opposite (attractive) sign of the coupling.
The critical properties of the Gross-Neveu and equivalent models have been studied extensively
in the literature 
[see, e.g., Refs.~\cite{Wetzel:twoloop,Ludwig:twoloop,Rossi2,Gracey:GrossNeveu,TracasVlachos,Rossi1,Kivel:GrossNeveu,Ludwig:Thirring,Gracey:FourLoop}],
with the critical exponents known up to fourth-loop order.
A similar two-loop RG analysis for graphene have been carried out in Ref.~\cite{Ostrovsky:grapheneRG}.
Critical exponents for Weyl particles in $2+\varepsilon$ dimensions have been evaluated explicitly to
two-loop order in Ref.~\cite{Syzranov:twoloopZ}.

Critical exponents for 3D Weyl and Dirac systems have been obtained numerically, using the critical scaling of the
DoS in Refs.~\cite{Herbut,Pixley:twotrans,Pixley:ExactZpubliahed,LiuOhtsuki:LateNumerics,Bera:Weyl}
and using conductance scaling in Ref.~\cite{Brouwer:exponents}; some of the results
are shown in Table~\ref{Table:NumericalExponents} and demonstrate that
the dynamical exponent is very close to the one-loop RG result $z=3/2$,
{while the correlation-length exponent may significantly deviate from the one-loop prediction $\nu=1$.}

\begin{table}[h!]
\begin{tabular}{c|c|c|c|c|c}
Ref. & \cite{Brouwer:exponents} & \cite{Herbut}  & \cite{LiuOhtsuki:LateNumerics} & \cite{PixleyHuse:rare2} & \cite{Bera:Weyl}
\\ \hline\hline
System & Single-node WSM                 & DSM            & WSM                            & WSM  				            & WSM
\\ \hline
$\nu$ & $1.47 \pm 0.03$         & $1.00\pm 0.15$ & $0.84\pm 0.1$                  & $1.01\pm0.06$                     & $0.72\pm 0.2 - 1.1\pm 0.15$
\\ \hline
$z$   & $1.49 \pm 0.02$         & $1.5\pm 0.1$   & $1.53\pm 0.03$                 & $z=1.50\pm 0.04$              & $1.38-1.49\pm 0.05$
\\ \hline

\end{tabular}
\caption{
\label{Table:NumericalExponents}
Numerical values of the correlation-length ($\nu$) and dynamical ($z$) exponents
for 3D Weyl and Dirac systems.}
\end{table}

\subsection{Density of states}
\label{Sec:DoS}

The existence of the critical point together with the assumption of the single-parameter scaling suggests
that the DoS near the critical point has the scaling form\cite{Herbut}
\begin{align}
	\rho(E,\varkappa)=(E-E_c)^{\frac{d}{z}-1}
	\Phi\left[(\varkappa-\varkappa_c)/(E-E_c)^\frac{1}{z\nu}\right]
	+\rho_{\text{smooth}},
	\label{CriticalDoS}
\end{align}
where $\Phi$ is an arbitrary function (that may be different for
$E<E_c$ and $E>E_c$). Here $\rho_{\text{smooth}}$ is a smooth analytic contribution,
which is generically allowed by the scaling theory and has been demonstrated microscopically 
to exist in models with smooth disorder\cite{Wegner:DoS,Suslov:rare,Nandkishore:rare,Skinner:WeylImp,Syzranov:unconv}.
The form (\ref{CriticalDoS}) of the critical DoS was first proposed (without
the smooth contribution) and verified numerically for 3D Dirac particles in Ref.~\cite{Herbut}.

\begin{figure}[htbp]
	\centering
	\includegraphics[width=0.5\textwidth]{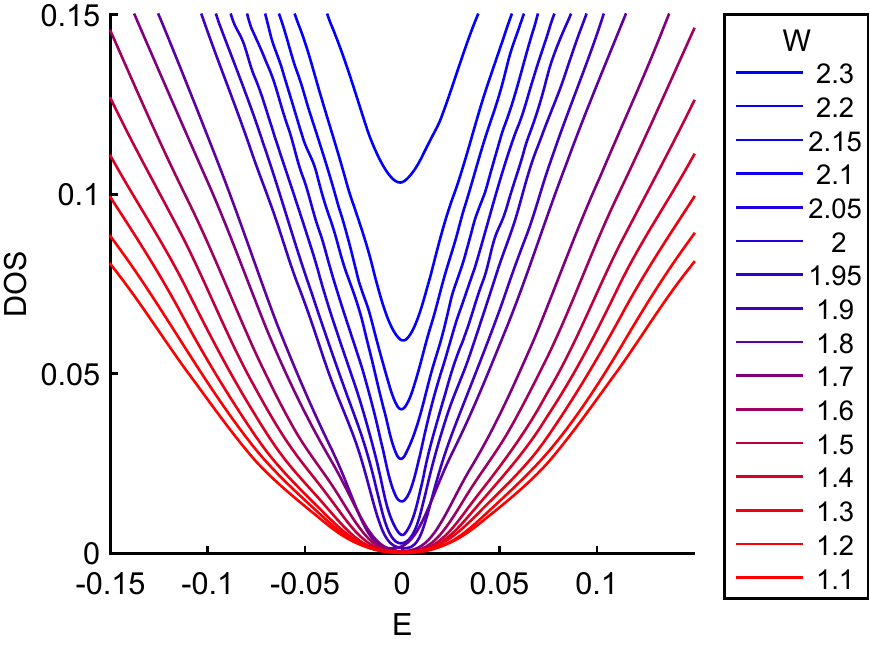}
	\caption{
	[Adapted from Ref.~\cite{LiuOhtsuki:LateNumerics}] Critical behaviour
	of the density of states near the non-Anderson transition in a Weyl semimetal
	as a function of energy for various random
	potential amplitudes $W$ ($\sim \varkappa^\frac{1}{2}r_0^{-\frac{d}{2}}$).
	The DoS near the node is strongly suppressed and can be approximated as
	$\rho\propto E^2$ if
	the disorder strength is smaller than a critical value\cite{LiuOhtsuki:LateNumerics}, $W<W_c\approx 2$.
	For $W>W_c$ the zero-energy density of states grows rapidly with the disorder amplitude $W$.
	}
	\label{DoScritical}
\end{figure}

Depending on the range of the disorder strength $\varkappa$ and energy $E$,
a high-dimensional system may exhibit several qualitatively different types of the DoS behaviour,
shown in Fig.~~\ref{PhaseDiagrSemicond},
which were first discussed for 3D Dirac systems in Ref.~\cite{Herbut}.
Near the critical disorder strength, $\varkappa\approx\varkappa_c$, at the ``critical regions''
in Fig.~\ref{PhaseDiagrSemicond}, the DoS behaves as $\rho\propto|E-E_c|^{\frac{d}{z}-1}
+\rho_{\text{smooth}}$. 
For strong disorder, $\varkappa>\varkappa_c$, the DoS is constant at low energies,
$\rho\propto(\varkappa-\varkappa_c)^{(d-z)\nu}+\rho_{\text{smooth}}$. 
For subcritical disorder, $\varkappa<\varkappa_c$,
the DoS has the same energy dependence as in a disorder-free system,
$\rho\propto(\varkappa_c-\varkappa)^{-d\nu\left(\frac{z}{\alpha}-1\right)}E^\frac{d-\alpha}{\alpha}+\rho_{\text{smooth}}$,
with the effects of disorder reflected in a prefactor that diverges at the critical point.
The scaling form (\ref{CriticalDoS}) of the DoS near nodes and band edges in semiconductors
and semimetals is supported by rigorous microscopic calculations \cite{Syzranov:unconv}.
The DoS in a 3D Dirac semimetal as a function of energy $E$ for various disorder strengths\cite{Herbut} is
shown in Fig.~\ref{DoScritical}.

We note that the analytic contribution $\rho_{\text{smooth}}$ is normally rather small
and has remained undetectable in all numerical studies\cite{Herbut,Brouwer:WSMcond,Garttner:longrange,
Pixley:twotrans,Pixley:ExactZpubliahed,LiuOhtsuki:LateNumerics,Brouwer:exponents,
Shapourian:PhaseDiagr,Bera:Weyl} so far, other than
Refs.~\cite{PixleyHuse:rare} and \cite{PixleyHuse:rare2}. This contribution
is determined by the rare-region effects discussed in Sec.~\ref{Sec:rare}.
It has also been suggested [see, e.g., Refs.~\cite{PixleyHuse:rare} and \cite{PixleyHuse:rare2}] that
such effects might smear the criticality in a small
region close to the transition, converting it into a sharp crossover. However, in most of this review
we do not distinguish between such a crossover and a real transition; more details on this topic
are presented in Sec.~\ref{Sec:rare}.


\subsection{Mobility threshold}

In a disordered system, localised and delocalised states do not coexist at the same energy\cite{Mott:edgeReview},
which allows one to introduce the concept of mobility threshold, the energy separating
localised and delocalised states\cite{Mott:reviewFirstMobThresh}.

The high-dimensional disorder-driven transition manifests itself in the unusual behaviour of the mobility
threshold as a function of disorder strength\cite{Syzranov:unconv},
provided the existence of both localised and delocalised states is allowed by dimensions, symmetry,
and topology, as in the case of $d>2$-dimensional semiconductors in the orthogonal symmetry class.

For weak disorder, $\varkappa<\varkappa_c$ the mobility threshold is pinned to the band edge due to the
smallness of the renormalised disorder strength. For supercritical disorder, however, the 
effective disorder strength increases near the band edge, until it reaches the strong-coupling point, $\gamma\sim 1$,
resulting in the localisation of all states 
in a trivial-topology system in the orthogonal symmetry class for $\gamma\lesssim d-2$ \cite{Efetov:book}.
Just above two dimensions this leads to the non-analytical behaviour of the mobility threshold\cite{Syzranov:unconv}
illustrated
in Fig.~\ref{PhaseDiagrSemicond}a,
which can be expected to hold in all high dimensions $d>\max(2\alpha,2)$.
This non-analytic behaviour of the mobility threshold should be contrasted with the
case of low dimensions $d<2\alpha$, where the mobility threshold is a smooth function of
the disorder strength\cite{Bulka,Efetov:book,GarciaGarcia}.

We note that for non-trivial topology and various disorder symmetries the phase diagram of a high-dimensional
disordered system can be more complicated. For example,
it has been shown recently\cite{Pixley:twotrans}
that the states near the nodes of a
3D Dirac system with axial and potential disorder experience two distinct transitions when
disordered strength is increased: the high-dimensional non-Anderson transition and, at larger disorder, the
conventional Anderson transition. The behaviour of the mobility threshold and its interplay with the non-Anderson
disorder-driven transition for Dirac (Weyl) systems can be studied rigorously analytically in $d=2+\varepsilon$
dimensions but currently remains a future research direction.

\subsection{Conductivity}
\label{Sec:cond}

The  unconventional
disorder-driven transition can be studied through the conductivity of a system, which is a quantity
readily
measured in condensed-matter experiments.
We note that in $d\leq 2$-dimensional systems in the orthogonal class with trivial topology,
such as arrays of trapped ions\cite{Monroe:longrange,Islam:longrange,Bollinger:longrange,Blatt:chain1,Blatt:chain2},
all excitations are localised, resulting in a vanishing conductivity, $\sigma=0$. 
By contrast, some systems, such as
chiral 1D chains\cite{Garttner:longrange},
may lack backscattering, in which case the conductivity is always nonzero.

Single-parameter scaling suggests a generic form of the conductivity\cite{Syzranov:Weyl}
(the response of the current of particles
at energy $E$ to an electric field),
\begin{align}
	\sigma(E,\varkappa)
	&\sim
	\xi^{-(d-2)}G[(E-E_c)\xi^{z}]
	\nonumber\\
	&=|\varkappa-\varkappa_c|^{\nu(d-2)}G[(E-E_c)/|\varkappa-\varkappa_c|^{z\nu}],
	\label{CondScaling}
\end{align}
where the scaling function $G$ is in general different for subcritical and supercritical disorder.

It follows from Eq.~(\ref{CondScaling})
that the conductivity at the node of a semimetal at zero temperature $T=0$ behaves
as $\sigma\propto |\varkappa-\varkappa_c|^{\nu(d-2)}$ \cite{Syzranov:Weyl}.
We note that the conductivity can be nonzero or vanish
on the strong-disorder side $\varkappa>\varkappa_c$, depending on the existence and the behaviour of the mobility
threshold. Equation~\eqref{CondScaling} also suggests that
in a disordered semimetal with the chemical potential at the nodal point, the conductivity
at the critical disorder strength 
($\varkappa=\varkappa_c$) exhibits the universal temperature dependence
$\sigma\propto T^{(d-2)/z}$ \cite{Syzranov:Weyl}.

\subsubsection*{Conductivity of a Weyl semimetal}

The conductivity near the unconventional disorder-driven transition has received particular attention
in the context of WSMs, which are currently the most experimentally available system in high
dimensions ($d>2\alpha$).

A remarkable feature of WSMs is the absence of localisation by smooth disorder
that does not scatter between different Weyl nodes. Because
a single-node WSM is the surface of a 4D topological insulator, it is topologically
protected from localisation\cite{RyuLudwig:classification,Wan:WeylProp}.
We note that realistic materials always have an even number of Weyl nodes, according to the fermion
doubling theorem\cite{NielsenNinomiya}. However, for sufficiently smooth disorder internodal
scattering can be neglected (e.g., if the length of internodal scattering exceeds the dephasing or magnetic length),
and the material behaves as an even number of independent copies of a single-node WSM. The conductivity in such system
is finite at all disorder strengths, with the possible exception of the critical point.

{For weak disorder, $\varkappa\ll\varkappa_c$, the conductivity of a single-node WSM
can be computed, e.g., using
the kinetic equation\cite{OminatoKoshino} or diagramatically\cite{Syzranov:Weyl,Lu:WSMcond},
with the result
\begin{align}
	\sigma=\frac{e^2v^2}{2\pi\varkappa}.
	\label{WSMcond}
\end{align}
Expressions similar to Eq.~\eqref{WSMcond}, but with different prefactors,
have been obtained also in Refs.~\cite{BurkovHookBalents} and \cite{HosurVishwanath}
neglecting the renormalisation of the velocity vertex by disorder.}

{Perturbative calculations\cite{OminatoKoshino,Syzranov:Weyl,Lu:WSMcond,BurkovHookBalents,HosurVishwanath}
of the conductivity, Eq.~(\ref{WSMcond}), valid for $\varkappa\ll\varkappa_c$, neglect the pseudoballistic
renormalisation of the system properties, which 
can be taken into account by means of the RG described in Section~\ref{Sec:RG}, with the infrared momentum cutoff determined by $E_F$ in an infinite system at zero temperature. This approach is valid below the
transition, $\varkappa<\varkappa_c$ (when disorder is irrelevant), but
is limited to high Fermi energies above the transition, 
$\varkappa>\varkappa_c$ [when disorder is relevant but must remain weak ($k_F\ell(k_F)\gg1$)].
The conductivity is given\cite{Syzranov:Weyl} by Eq.~(\ref{WSMcond})
with the ``bare'' disorder strength $\varkappa$ replaced by its renormalised value $\tilde\varkappa(E_F)$
[for the SCBA analysis of the conductivity see also Ref.~\cite{OminatoKoshino}].}

\begin{figure}[ht]
	\centering
	\includegraphics[width=0.55\textwidth]{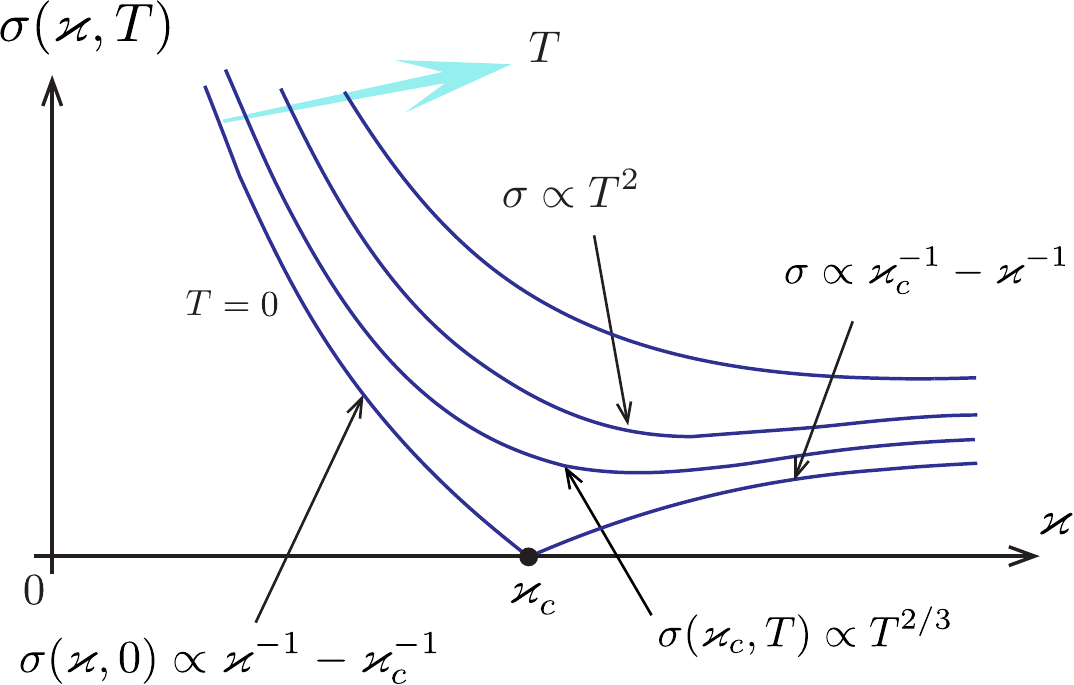}
	\caption{
	[Adapted from Ref.~\cite{Syzranov:Weyl}] Conductivity of a single-node WSM 
	as a function of disorder strength for various temperatures.
	The exponents for the dependencies on disorder strength and temperature
	near the transition are obtained within perturbative one-loop RG treatment.
	}
	\label{WSMCondCrit}
\end{figure}

For strong disorder, $\varkappa>\varkappa_c$, the conductivity has also been obtained using order-of-magnitude
estimates and the scaling theory in Ref.~\cite{Syzranov:Weyl}. 
An effective field theory for WSMs at strong disorder has been developed in Refs.~\cite{BagretsAltland:smodel1}
and \cite{BagretsAltland:smodel2}. The conductivity of a single-node WSM as a function of the disorder
strength $\varkappa$ and temperature $T$ is shown in Fig.~\ref{WSMCondCrit}.

{Transport in WSMs has been studied numerically in Refs.~\cite{Brouwer:WSMcond}
and \cite{Brouwer:VectorWeyl}, utilising a smooth random potential with a long correlation length resulting
in suppressed internodal scattering. References~\cite{Brouwer:WSMcond}
and \cite{Brouwer:VectorWeyl} reported a vanishing conductivity for subcritical disorder $\varkappa<\varkappa_c$
and a finite conductivity for supercritical disorder $\varkappa>\varkappa_c$.
Such a behaviour seemingly contradicts Eq.~\eqref{WSMcond}, the result of rigorous analytical 
calculations\cite{OminatoKoshino,Syzranov:Weyl,Lu:WSMcond,BurkovHookBalents,HosurVishwanath},
which gives a nonzero conductivity for $\varkappa\ll\varkappa_c$.
As noted in Ref.~\cite{Brouwer:WSMcond}, this discrepancy is explained by the noncommutativity of the
limits $E_F=0$, $L\rightarrow\infty$ and $E_F\rightarrow0$, $L=\infty$, where the Fermi energy $E_F$
is measured from the Weyl node and $L$ is the system size. Numerical simulations\cite{Brouwer:WSMcond,Brouwer:VectorWeyl} carried out for finite samples and zero chemical potentials yield zero weak-disorder conductivity (extracted
respectively from the conductance), while analytical calculations\cite{OminatoKoshino,Syzranov:Weyl,Lu:WSMcond,BurkovHookBalents,HosurVishwanath} deal with the limit of an infinite sample
and the chemical potential sent to zero at the end of the calculation, which results in a finite conductivity.
This non-commutativity is explained by finite-size effects\cite{Brouwer:WSMcond,Syzranov:Weyl}
discussed in detail in Sec.~\ref{Sec:FiniteSize}.}

{We emphasise that the 
calculations~\cite{OminatoKoshino,Syzranov:Weyl,Lu:WSMcond,BurkovHookBalents,HosurVishwanath,BagretsAltland:smodel1}
of the conductivity in a WSM, discussed in this section, are based on
the assumption that the disorder potential may be considered short-range [cf. Eq.~(\ref{DisorderShortStrength})]
for scattering processes within one node.
By contrast, in solid-state WSMs disorder is dominated normally by Coulomb impurities,
leading to long-range random potential. The presence of charged impurities may result in a finite conductivity
for all disorder strengths\cite{Skinner:WeylImp}, as discussed in Sec.~\ref{Sec:ExpWSMsolid} below.
}

\subsection{Multifractality}

Wavefunction multifractality (fractal structure characterised by an infinite set of exponents)
has been a subject of extensive studies in
the context of Anderson and quantum Hall transitions [see Refs.~\cite{Efetov:book,MirlinEvers}
for review] and is characterised by the inverse participation ratios\cite{Wegner:IPR} (IPRs) $P_q=\int |\psi(\br)|^{2q}d\br$.
{
These quantities describe how extended the wavefunction $\psi(\br)$ is and have two obvious limits: 
for delocalised states that uniformly fill the whole system $P_q\propto L^{-d(q-1)}$,
while for localised states the IPRs are system-size-independent, $P_q=\text{const}$.}
For multifractal wavefunctions disorder-averaged IPRs behave as $P_q\propto L^{-d(q-1)-\Delta_q}$
with a non-linear (as a function of $q$) multifractal spectrum $\Delta_q$ (where $\Delta_q=0$ corresponds to
the extended plane-wave-like states $\psi\propto L^{-\frac{d}{2}}$).

It has been shown recently that the high-dimensional disorder-driven transition
also exhibits wavefunction multifractality
(regardless of whether the system allows for a localisation transition),
with the spectrum different from that of the Anderson transition.
{Within an $\varepsilon=d-2\alpha$-expansion the multifractal spectrum is given by\cite{Syzranov:multifract}
\begin{align}
	\Delta_q=\frac{1}{2}(2\alpha-d) q(q-1)+\cO[(d-2\alpha)^2]
	\label{FractSemicond}
\end{align}
for a semiconductor in the orthogonal symmetry class and
\begin{align}
	\Delta_q=-\frac{3}{8}(d-2)^2 q(q-1)+\cO[(d-2)^3]
	\label{FractWeyl}
\end{align}
for Weyl particles in $d$ dimensions.} 
The multifractal spectrum \eqref{FractWeyl} matches that
of 2D Dirac fermions (with $\varepsilon$ replaced by the dimensionless disorder strength), as studied
in Ref.~\cite{Foster:2Dfractality}. We note that for 2D Dirac fermions there is no phase transition, although
on sufficiently short length scales the wavefunctions display {\it non-universal} fractal behaviour\cite{Foster:2Dfractality,ChouFoster:NumerFractality}.
By contrast, the multifractal behaviour described by Eqs.~\eqref{FractSemicond} and \eqref{FractWeyl} persists
on all length scales at non-Anderson disorder-driven transitions.
{As a result of multifractality, moments of physical observables (e.g. conductance or DoS)
may be expected to have infinite sets of critical exponents at the transition point.
Also, different scaling behaviour of typical and average DoS, observed recently
in a Dirac semimetal in Ref.~\cite{Pixley:twotrans}, has been argued to be a reflection of
of the wavefunction fractality\cite{Pixley:twotrans}.}

\section{FINITE-SIZE EFFECTS IN HIGH DIMENSIONS}
\label{Sec:FiniteSize}

Finite-size effects in high dimension $d>2\alpha$ 
are qualitatively different from the case of low dimensions
and lead to several qualitatively new phenomena.

\subsection{Level discreteness and ballistic transport in disordered systems}

{A finite disorder-free system is characterised by a discrete spectrum of energy levels, $E_n$, 
as a result of spatial quantisation. In the presence of disorder, the levels get broadened, which
results in a smooth disorder-averaged DoS $\rho(E)$ as a function of energy 
if the elastic scattering rate $\tau^{-1}(E)$
exceeds the level spacing $|E_n-E_{n-1}|$. 
In a conventional (low-dimensional, $d<2\alpha$) system the disorder-averaged 
DoS $\rho(E)$ in the presence of arbitrarily
weak disorder is a smooth function for all energies $E$, which are allowed by the bandwidth and disorder amplitude,
in the limit of infinite system size.}

{A contrasting remarkable feature of high-dimensional ($d>2\alpha$) systems is that the levels
near nodes and band edges remain discrete\cite{Syzranov:multifract},
with a sharply peaked DoS $\rho(E)$ (see Fig.~\ref{DoSPeaks}),
for disorder $\varkappa<\varkappa_c$, irrespective of how large the system is
(and provided rare-region effects discussed in Sec.~\ref{Sec:rare} are neglected).
Specifically,
the level spacing $|E_n-E_{n+1}|$ for energy levels closest to
the node or a band edge exceeds the ``level width'' given by the elastic
scattering rate\cite{Syzranov:multifract,Syzranov:Weyl}.
This is a reflection of
the elastic scattering rate $\tau^{-1}(E)\sim\gamma(E)\,(E-E_c)$ decreasing faster than the energy $E-E_c$
when approaching a node or a band edge, as discussed above.
\begin{figure}[ht]
	\centering
	\includegraphics[width=0.5\textwidth]{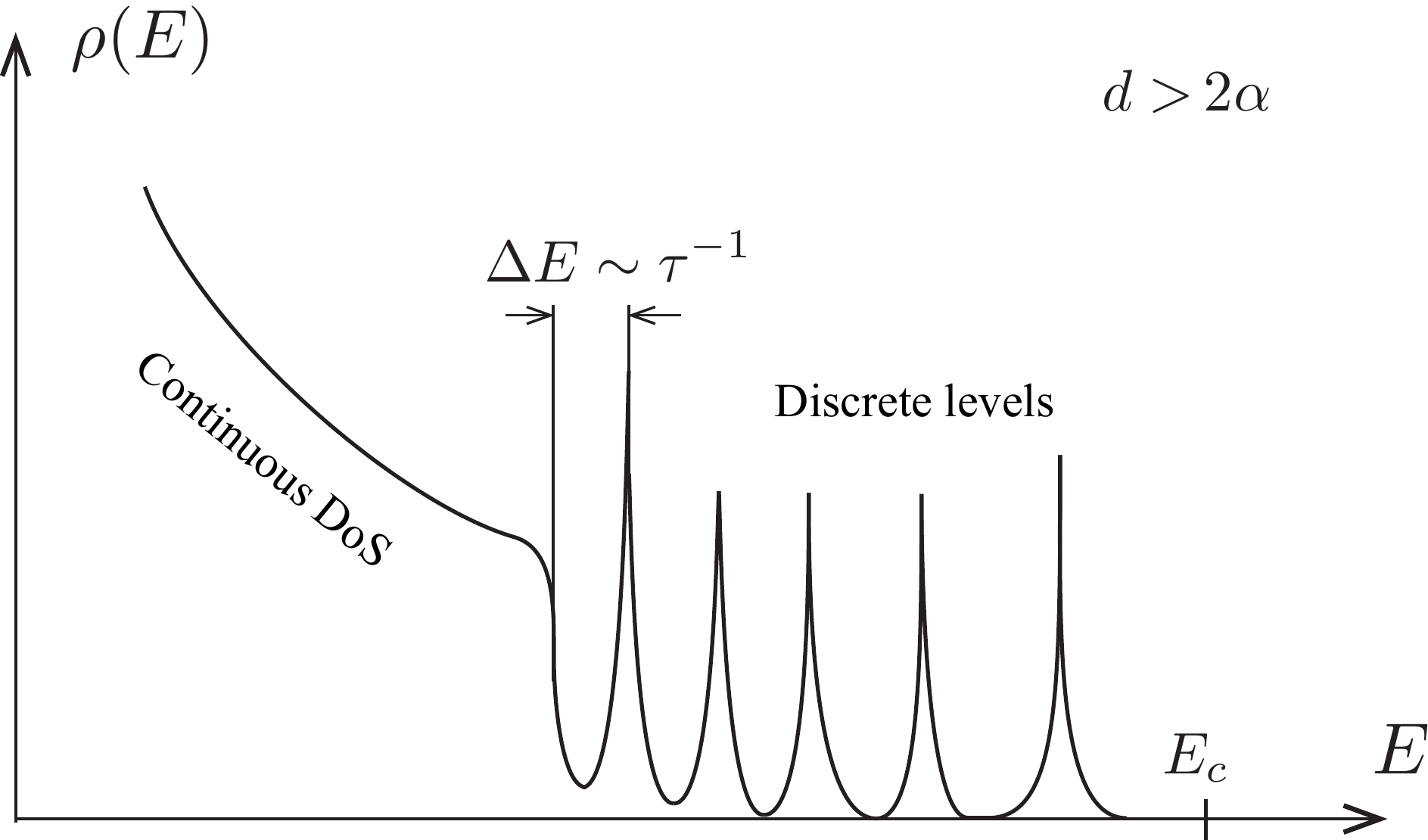}
	\caption{
	\label{DoSPeaks}
	The density of states near a band edge in a high-dimensional ($d>2\alpha$) system.
	At low energies the DoS necessarily exhibits level discreteness, which is a reflection
	of the irrelevance of subcritical disorder.
	}
\end{figure}

{When the system size is increased, the spacing between all levels decreases, but the level width of the
lowest-energy levels decreases faster than the spacing, so that the lowest levels always remain discrete.} 
}
For a system with the small parameter $|\varepsilon|=|d-2\alpha|\ll1$,
there are $N_\varepsilon\sim 1/|\varepsilon|$ of discrete levels (sharp peaks of the DoS) at the
critical disorder $\varkappa=\varkappa_c$, while the rest of the spectrum
is smooth\cite{Syzranov:multifract}.

Moreover, for the discrete levels the elastic scattering time $\tau$ exceeds the ballistic Thouless time
$\tau_{\text{Th}}=L/v$ required for a particle to propagate through the system. 
The transport of quasiparticles sufficiently close to band edges and nodes is
therefore always ballistic in arbitrarily large high-dimensional systems
for $\varkappa<\varkappa_c$.

\subsection{Conductance of finite semimetallic samples}

\begin{figure}[h!]
	\centering
	\includegraphics[width=0.9\textwidth]{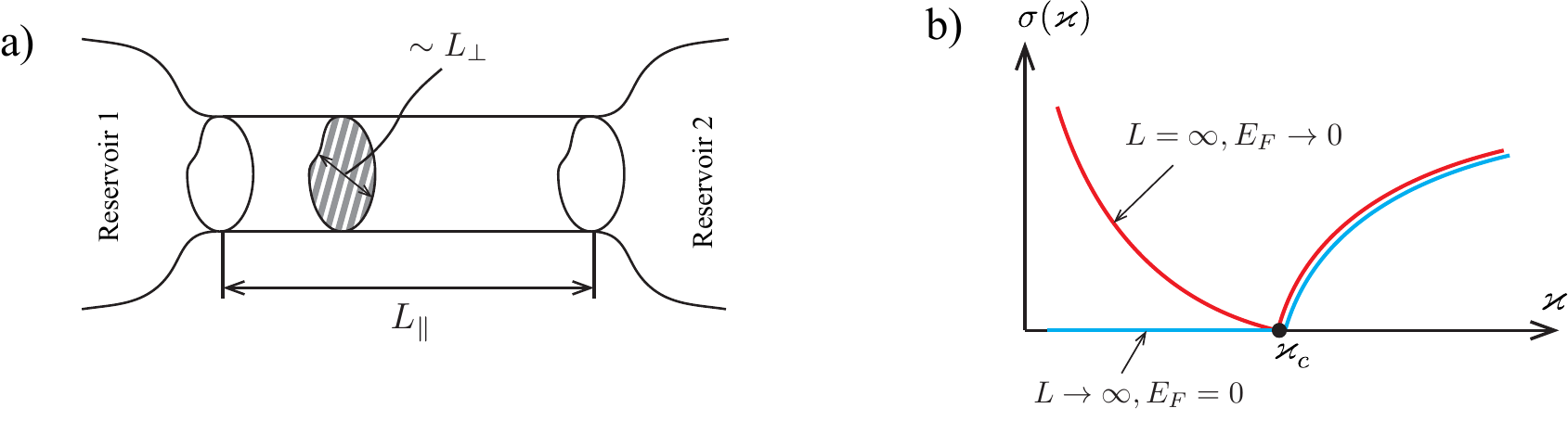}
	\caption{ a) Set-up for measuring the conductance of a finite sample.
	b) Conductivity of a single-node Weyl semimetal in the limits of an infinite sample and small finite Fermi energy
	($L=\infty$, $E_F\rightarrow 0$) and of a large finite sample and zero Fermi energy
	($L\rightarrow\infty$, $E_F=0$). The two limits do not commute
	for subcritical disorder, $\varkappa<\varkappa_c$ \cite{Brouwer:WSMcond,Syzranov:Weyl}.
	}
	\label{ConductanceSetup}
\end{figure}

The level discreteness discussed above is reflected in the conductance of a finite
semimetallic sample in high dimensions
if the Fermi energy $E_F$ (measured from the node) and the temperature are smaller than the level
spacing near the node\cite{Syzranov:Weyl}. 
Indeed, the transverse momentum is quantised in a finite sample between two reservoirs (Fig.~\ref{ConductanceSetup}a). Depending on the boundary conditions in the transverse direction,
a small Fermi energy $E_F$ either lies below the minimal excitation energy or corresponds to only one
transverse mode. The conductance of such a system, therefore, does not exceed the spin and valley degeneracy
of one transverse mode, $G\lesssim g$, according to the Landauer formula.

Therefore, the conductivity of a finite semimetallic sample, defined as $\sigma=G\frac{L_\parallel}{S_\bot}
\lesssim \frac{gL_\parallel}{L_\bot^2}$,
vanishes in the limit of infinite system size, $L_\parallel\propto L_\bot\rightarrow\infty$,
where $L_\bot$ and $L_\parallel$ are the transverse and longitudinal sizes of the sample, as shown
in Fig.~\ref{ConductanceSetup}a.
Vanishing conductivity for $\varkappa<\varkappa_c$ 
and Fermi level located at the node ($E_F=0$)
has been reported in numerical simulations of
transport in finite-size WSM systems in Refs.~\cite{Brouwer:WSMcond} and \cite{Brouwer:VectorWeyl},
in accordance with the above argument.

In contrast, analytical calculations of conductivity\cite{OminatoKoshino,Syzranov:Weyl,Lu:WSMcond,BurkovHookBalents,HosurVishwanath}, 
discussed in Sec.~\ref{Sec:cond} and based on the kinetic equation and the
Kubo formula, deal with the limit
of an infinite sample, $L_\parallel=L_\bot=\infty$, and
a finite chemical potential (that may be taken to zero at the end of the
calculation), leading to a nonzero conductivity of a WSM for subcritical disorder, $\varkappa<\varkappa_c$.

The behaviour of the conductivity as a function of disorder strength 
is shown in both limits in Fig.~\ref{ConductanceSetup}b.
As noted in Ref.~\cite{Brouwer:WSMcond}, each limit can be effectively realised in experiment.
Finite conductivity (the infinite-sample regime) can be expected 
if the temperature or the rate of inelastic scattering
exceed the level spacing near the node. {Otherwise, the limit of a finite system and zero Fermi energy
is realised, giving zero conductivity for subcritical disorder strengths.}

We note that, in addition to the bulk states discussed above, a material can host surface states
that also contribute to conductance. One important example is the contribution
of the surface states of a WSM\cite{Wan:WeylProp} to the conductance, which
can be estimated as\cite{LiuOhtsuki:LateNumerics}
$G_{\text{surf}}\sim L_\bot K_0$, with $K_0$ being the separation between the Weyl nodes in momentum space.
This surface component
significantly exceeds the bulk contribution to the conductance of an undoped WSM at low temperatures.

\section{RARE-REGION EFFECTS}

\label{Sec:rare}

Perturbative RG and SCBA analyses in high-dimensional systems suggest that the DoS
vanishes at nodes and band edges for subcritical disorder and has a finite
value otherwise. {Thus, the RG and SCBA approaches} suggest that the DoS at nodes and band edges
can serve as an order parameter for the unconventional disorder-driven transition.

However, it has been known for three decades\cite{Wegner:DoS} that disorder with a smooth distribution function
always leads to a non-vanishing disorder-averaged density of states at all energies allowed 
by the band width and the amplitude of the random potential. Strictly speaking,
nodes (points with zero DoS away from band edges) do not exist in disordered systems,
and band edges exist only for bounded (having finite maximal amplitude)
disorder. 
Concomitantly, this well-known result demonstrated that the DoS is not an order parameter that
can distinguish between low- and high-disorder phases [as was emphasised recently for Weyl semimetals
in Ref.~\cite{Nandkishore:rare}].

This broadening of nodes and the appearance of states far below the band edges in disorder-free samples
occurs as a result of statistically rare non-perturbative fluctuations of the disorder potential,
that is
unaccounted for by the perturbative RG.

We emphasise, however, that the DoS not being an order parameter (or even the absence of any order parameter)
does not preclude the existence of a genuine sharp transition manifested, for example, in non-analytic
behaviour of physical observables across the critical point (as exemplified by Kosterlitz-Thouless transition).

\subsection{Lifshitz tails}

Rare fluctuations of the disorder potential lead to the existence of arbitrarily
deep potential wells that trap particle states (provided the existence of localised states
is allowed by symmetry and/or topology), leading to the formation of ``Lifshitz tails'',
states with energies significantly below the band edge of a disorder-free system,
as shown in Fig.~\ref{PhaseDiagrSemicond}.
Lifshitz tails have been
studied extensively\cite{Lifshitz:tail,ZittartzLanger,HalperinLax,LifshitzGredeskulPastur} for
conventional low-dimensional ($d<2\alpha$) semiconductors. 
The DoS $\rho_{\text{tail}}$ in Lifshitz tails is
non-universal and depends strongly on the microscopic details
of disorder.
For Gaussian disorder and quadratic quasiparticle dispersion, $\rho_{\text{tail}}(E)\propto\exp(-c|E|^{2-d/2})$
for sufficiently low energies $E$ and $d<4$, with $c$ being a constant. In the field-theoretical representation
[cf. action (\ref{Action})]
the rare-region effects correspond to instantons [see Ref.~\cite{Yaida:instantons} for review],
which cannot be taken into account at any order
in perturbation theory.

In high dimensions, the form of Lifshitz tails is qualitatively different from the case of low-dimensions.
It has been shown in Ref.~\cite{Suslov:rare}, by analysing instantonic contributions to the replica field theory,
that for $d>4$ and quadratic dispersion, the DoS in the
presence of Gaussian disorder of strength $\varkappa$ behaves according to
\begin{align}
	\rho_{\text{tail}}(E)\propto e^{-c_1(c_2+|E|)^2/\varkappa},
	\label{TailDoS}
\end{align}
where $c_1$ and $c_2$ are $\varkappa$-independent constants and the disorder correlation length is assumed
fixed.
In Ref.~\cite{Syzranov:unconv} it was shown that the dependency (\ref{TailDoS}) of the DoS in Lifshitz tails on energy
and disorder strength
persists generically for systems with the dispersion $\epsilon_\bk\propto k^\alpha$ in any dimension $d>2\alpha$.

We note that in semimetals that allow for localisation
the smearing of nodes can be also effectively viewed as the beginning of Lifshitz tails.
However, in semimetals that are topologically protected from localisation the smearing of nodes
requires a separate analysis. In particular, a detailed microscopic analysis
of the DoS for a single-node Weyl semimetal has been carried out recently
 in Ref.~\cite{Nandkishore:rare}. It has been demonstrated 
that the DoS at the node is finite and   
exponentially small
in disorder strength.

In systems just above the critical dimension ($1\gg d-2\alpha>0$), the DoS in the Lifshitz tail is strongly
suppressed at all energies, 
with $\rho_{\text{tail}}(0)\propto e^{-\frac{1}{d-2\alpha}\frac{\varkappa_c}{\varkappa}}$ \cite{Syzranov:unconv},
which near the transition ($\varkappa\lesssim\varkappa_c$) allows one to separate accurately the band states
(``weak-disorder'' phase in Fig.~\ref{PhaseDiagrSemicond}) from the tail.
This contrasts with the case of low dimensions ($d<2\alpha$),
where the beginning of the Lifshitz tail is not parametrically suppressed. 
One may show that a large number of valleys $N$ also leads to the suppression of the
rare-region contribution to the DoS, with $\rho_{\text{tail}}(0)\propto e^{-{N}\frac{\varkappa_c}{\varkappa}}$.
Thus, the contributions of the rare-region effects to the density of states are suppressed when the
system properties are computed in large-$N$ and small-$\varepsilon$-controlled RG approaches. This allows
one to use the DoS as an effective order parameter for the unconventional disorder-driven transition
so long as the exponentially small rare-region contribution is neglected. 

Realistic 3D Dirac and Weyl semimetals ($\alpha=1$, $d=3$), the best known and most extensively
studied class of systems for the high-dimensional phenomena discussed in this review,
have neither a small parameter $\varepsilon$ nor a large number of valleys $N$. Nevertheless, the
contributions of rare-region effects to physical observables in these systems appear to be rather small
and have been undetectable in most numerical studies of the DoS and
transport\cite{Herbut,Brouwer:WSMcond,
Pixley:twotrans,Pixley:ExactZpubliahed,LiuOhtsuki:LateNumerics,Brouwer:exponents,
Shapourian:PhaseDiagr,Bera:Weyl} to date, with the exception of the recent simulations \cite{PixleyHuse:rare,PixleyHuse:rare2}.
Because such systems have no small parameters at the critical point, the reason for the suppression
of rare-region effects in 3D Weyl (Dirac) systems, observed numerically, currently remains to be investigated.

\subsection{Phase transition vs. sharp crossover}

Another potential consequence of rare-region effects, currently not demonstrated
analytically, is a possible broadening of the quantum critical region,
which may convert the transition into a sharp crossover.
So far we have not distinguished between these two scenarios and we have used the term
``non-Anderson disorder-driven transition'' to label either a true transition or a sharp crossover.

The elimination of the transition has been advocated recently in the numerical studies of 3D Dirac semimetals
in Refs.~\cite{PixleyHuse:rare} and \cite{PixleyHuse:rare2}, which suggest smooth and analytic behaviour
of the disorder-averaged DoS, which in turn was additionally averaged with respect to the boundary conditions
in a finite sample.

We note that the phenomenology of rare-region effects and their interplay with perturbative effects
are qualitatively distinct in systems where Anderson localisation is allowed by symmetry and topology
and systems where localisation is forbidden. While in the conventional analyses\cite{Lifshitz:tail,ZittartzLanger,HalperinLax,LifshitzGredeskulPastur,Suslov:rare} rare-region effects come
from localised states, their nature should be different in materials like single-node WSMs
and 1D chiral chains\cite{Garttner:longrange}, where states are topologically delocalised,
and thus deserve a separate study.

While in a WSM power-law-localised (bound) states $\psi\sim 1/r^2$ exist in a single potential well\cite{Nandkishore:rare},
in a random potential such states (decaying slower than $1/r^d$) inevitably get hybridised with each other
(as well as with other extended states\cite{Mott:reviewFirstMobThresh,Mott:edgeReview})
and form extended states, according to the renormalisation-group procedure developed
in Refs.~\cite{Levitov:AnnReview,Levitov1,Levitov2} [see also Ref.~\cite{Nandkishore:rare}].
A Weyl semimetal in the presence of a random gauge field is another example of a system
where the conventional analysis of rare regions cannot be applied\cite{Brouwer:VectorWeyl} and awaits further studies.

\section{POTENTIAL FOR EXPERIMENTAL REALISATION}

Despite extensive theoretical studies, non-Anderson disorder-driven transitions
still remain to be observed experimentally.
In what follows we consider various systems that have the potential for observing such transitions.

\subsection{Weyl and 3D Dirac semimetals}

3D Weyl and Dirac semimetals ($\alpha=1$) are currently among the most
experimentally accessible systems in the
high-dimensional regime ($3=d>2\alpha=2$). 

\subsubsection{Solid-state semimetals} 
\label{Sec:ExpWSMsolid}

Recently 3D Dirac quasiparticles have been found experimentally in $Na_3Bi$ \cite{Liu:Na3BiDiracdiscovery}
and $Cd_3As_2$ \cite{Liu:Cd3As2Diracdiscovery}. WSMs have been discovered 
in $TaAs$ \cite{ZHasan:TaAs,ZHasan:TaAs2,Weng:PhotCrystWSM}, $TaP$ \cite{ZHasan:TaP} and $NbAs$ \cite{ZHasan:NbAs}.

While there are many types of disorder in solid-state systems, transport in realistic materials is typically
dominated by charged impurities.
In Weyl and Dirac semimetals,
the disorder correlation length is given by the impurity screening length,
$\lambda\sim v/[\alpha^\frac{1}{2}\max(\mu,T)]$,
where
$\alpha=\frac{e^2}{v\kappa}$ is the ``fine-structure constant''
and $\mu$ is the characteristic chemical potential [determined at low doping by the fluctuations
of the impurity concentration\cite{Skinner:WeylImp}].
Because in realistic WSM materials $\alpha\lesssim 1$, this screening length
exceeds the characteristic electron wavelength $\lambda_e\sim v/\max(\mu,T)$ 
that determines the WSM conductivity. Because the observation of the non-Anderson disorder-driven transitions 
requires that the random potential be short-range (compared to the electron wavelength),
these transitions cannot be observed
in transport\cite{Skinner:WeylImp,BurkovHookBalents,Ominato:Coulomb,HwangSarma,Lundgren:thermoelectric,Adam:magnetoresistance,Rodionov:donorsacceptors,PesinLevchenko:DoS,KlierMirlin:magnetoresistance}.

However, even in the presence of Coulomb impurities 
the transition can be observed in the DoS of doped systems\cite{Rodionov:donorsacceptors}.
Indeed, the momentum states near a WSM node below the Fermi energy
have large wavelengths that exceed the screening length $\lambda$ set by the Fermi energy. Although such states
do not contribute to conduction, they can be probed, for example in ARPES experiments, revealing the
critical behaviour of the DoS at the transition near the node, which was discussed in Sec.~\ref{Sec:DoS}.

\subsubsection{Weyl and Dirac semimetals in cold-atomic systems and photonic crystals}

Theoretical activity on Weyl semimetals has also spurred numerous proposals for realising Weyl 
and Dirac semimetals in cotrollable ultracold atomic 
gases [see, for example, Refs.~\cite{Lan:WSMprop,Jiang:WSMprop,Spielman:WSMprop,Soljacic:WSMprop,Ganeshan:WSMprop,Liu:SFimplementation,Lepori:PTlattice,Lepori:WSMrealisation}].
Moreover, it has been demonstrated\cite{Syzranov:WeylDipGases}
that excitations with Weyl dispersion exist in all 3D systems
of dipolar particles in the presence of magnetic field.
Although cold-atomic 3D Weyl (Dirac) semimetals still remain to be realised,
such systems are a promising platform for observing the physics of non-Anderson disorder-driven transitions.
In particular, such systems allow one to realise short-ranged disorder potentials
whose strength is readily controlled in experiment.

Recently a WSM has been realised\cite{Weng:PhotCrystWSM} in a photonic crystal,
a periodic structure with a broken inversion symmetry exposed to microwave radiation.  
Systems of such type allow for the controlled inclusion of short-range disorder and thus
present another flexible platform for studying high-dimensional non-Anderson disorder-driven transitions.

\subsection{1D and 2D arrays of ultracold trapped ions}

Arrays of trapped ultracold ions with long-range spin interactions $\propto 1/r^{d+\alpha}$ have recently 
been realised
in 1D\cite{Monroe:longrange,Islam:longrange,Blatt:chain1,Blatt:chain2}
and 2D\cite{Bollinger:longrange} lattices in magnetic traps, with 
a tunable parameter $\alpha$.
In a strong magnetic field, the propagation of single-spin-flip excitations
in these systems is effectively a single-particle problem, with the particle dispersion $\varepsilon_\bk\propto k^\alpha$.

\begin{figure}[h!]
	\centering
	\includegraphics[width=0.95\textwidth]{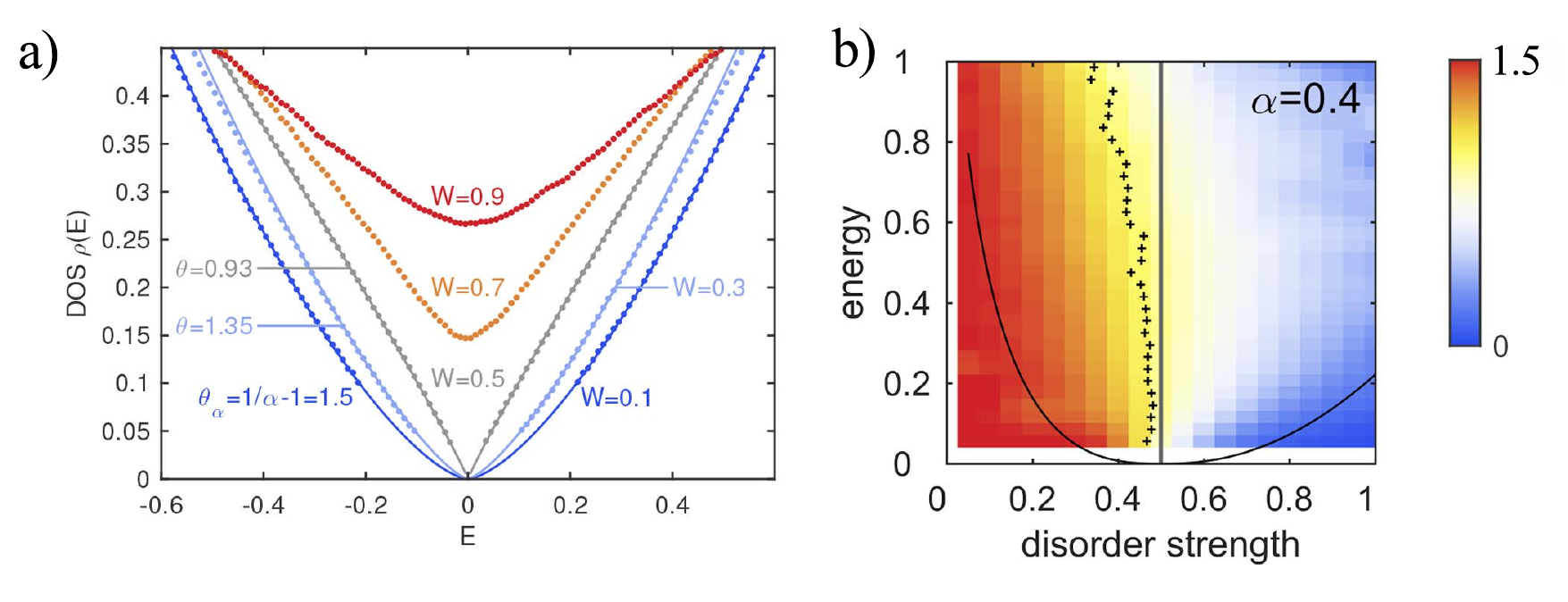}
	\caption{\label{ChiralChainPlots}
	[Adapted from Ref.~\cite{Garttner:longrange}]
	The critical behaviour of the density of states in a chiral 1D chain
	with the quasiparticle dispersion
	$\epsilon_k=|k|^\alpha\sign k$, calculated numerically in Ref.~\cite{Garttner:longrange}.
	a) DoS as a function of energy. For subcritical disorder the DoS vanishes $\propto E^{\frac{d}{\alpha}-1}$ near $E=0$.
	For supercritical disorder the low-energy DoS is nonzero. 
	At the critical disorder $\rho(E)\propto E^{\frac{d}{z}-1}$ b) Critical behaviour of the DoS
	on the diagram energy vs. disorder strength;
	the colour shows the exponent $\theta=\frac{\partial\ln\rho}{\partial\ln E}$.}
\end{figure}

When $\alpha$ is tuned to values $\alpha<d/2$, these systems correspond to the case of high dimensions.
When also subject to a weak disorder (e.g. a spatially random 
magnetic field), such systems present a flexible platform\cite{Garttner:longrange}
for observing non-Anderson disorder-driven transitions. 
{Similarly to the case of solid-state systems, arrays of trapped ions with $\alpha<d/2$ display
the critical behaviour of the DoS and other observables, as discussed in Sec.~\ref{Sec:observables}
[the numerical results\cite{Garttner:longrange} for the critical behaviour of the DoS in a 1D chiral chain are shown in Fig.~\ref{ChiralChainPlots}].
}

We note that current experiments with 1D chains\cite{Islam:longrange} use relatively small numbers ($N\leq 16$)
of ions, that may be insufficient for accurate observations of the transition.
An additional difficulty is the accurate detection of the finite energy $E_c$ of the critical point.
This difficulty, however, could be overcome in chiral systems\cite{Garttner:longrange} with the
quasiparticle dispersion $\epsilon_\bk\propto |k|^\alpha\sign k$, where the symmetry ensures $E_c=0$,
although such chiral systems still remain to be designed and realised.
It is likely that observing high-dimensional non-Anderson transitions will require longer chains of trapped ions
or using 2D arrays [which currently contain\cite{Bollinger:longrange} approximately $300$ ions].

\subsection{Chiral superconductors}

Another class of systems with the potential to realise high-dimensional disorder-driven transitions
are 3D nodal superconductors.
Such systems generically have points and/or lines on the Fermi surface where the superconducting gap vanishes.
They are thus described by Bogolyubov-de-Gennes Hamiltonians with nodal quasiparticles of the form
considered in this review,
e.g., chiral $p_x\pm ip_y$ superconductors that host Weyl excitations. 
Therefore, such systems can be expected to display the non-Anderson disorder-driven
transitions (provided superconductivity is not destroyed by disorder), that
can be detected, e.g., in the quasiparticle density of states.
These transitions in 3D chiral superconductors deserve further investigation.

At present, the existence of 3D chiral superconductors is not completely established,
but there is evidence for their existence in several candidate materials (see Ref.~\cite{Kallin:review}
for review).
A prominent example here is a the candidate $d$-wave superconductor $URu_2Si_2$ \cite{Yano:angleResolved,Kasahara:thermalcond,Kapitulnik:Kerr,
Fernandes:URuSi}.

\subsection{Quantum kicked rotors} 

The dynamics of a quantum rotor kicked by perturbations with $d$ incommensurate frequencies
can be mapped onto the problem of localisation of a quadratically dispersive particle
in $d$ dimensions in a (quasi-)random potential\cite{Grempel:mapping,Casati:mapping}.
{Thus, such systems may be used to study localisation phenomena and disorder-driven transitions,
both Anderson and non-Anderson, in various dimensions.}
{Quantum kicked rotors have been realised in systems of cold-atoms exposed to pulsed laser beams
to demonstrate
localisation in 1D systems\cite{Moor:rotorrealisation} and the Anderson localisation transition
in 3D\cite{Chabe:rotorrealisation,Delande:rotorrealisation}}.
These systems may be used similarly to demonstrate non-Anderson disorder-driven transitions
in $d>4$ dimensions and to study the properties of these transitions.

\subsection{Numerical simulations in high dimensions}

Numerical simulations allow access to all dimensions $d$, including those above the physical ($d=1,2,3$), and
thus can serve as another platform for the investigation of high-dimensional non-Anderson transitions.
{Often numerical simulations are carried out for models on lattices, which always
have finite bands and quadratic particle dispersion $\epsilon_\bk\propto k^2$
near band edges. 
Non-Anderson disorder-driven transitions can therefore be observed in such systems
in dimensions $d>4$ near band edges, however, so far the studies\cite{Markos:review,GarciaGarcia,Slevin,Zharekeshev:4D,SlevinOhtsuki:HighDUnitary} of high-dimensional localisation
phenomena focussed in states in the middle of the band and on the usual Anderson-transition
criticality.
Also, transport in disordered systems with arbitrary quasiparticle dispersion $\epsilon\propto k^\alpha$
may be simulated in momentum space using the scattering-matrix
approach\cite{Brouwer:WSMcond,Brouwer:exponents,Brouwer:VectorWeyl}, which allows one to
study non-Anderson disorder-driven
transitions for arbitrary $\alpha$ and $d>2\alpha$.}

\section{OPEN QUESTIONS AND FUTURE PROBLEMS}

Below we list questions that in our view constitute some
important further research directions related to the unconventional
high-dimensional disorder-driven physics.

\subsubsection*{Interactions, {magnetic field} and more general disorder and band structures.} 
A natural further research direction is the inclusion of electron-electron interactions and the study of its interplay with disorder. Recently such an interplay has been studied for Dirac semimetals in Refs.~\cite{Goswami:TIRG}
and \cite{Moon:RG}, however, a number of open questions remain, particularly, a more detailed analysis of physical observables. Also, most of the focus in the literature has been on potential disorder, while more generic disorder (random gauge field, random mass, etc.) is
expected\cite{LudwigFisher,AleinerEfetov} to appear. 

Furthermore, recent studies have demonstrated [see, e.g., Ref.~\cite{Xie:NLSM}] the existence of more general band structures, such as line nodes, some appearing in the context of chiral superconductors.
The role of generic disorder for a variety of nodal band structures in the presence of interactions
and, in particular, their effects on the unconventional disorder-driven transition
remains an open problem. {Another future direction of research is the interplay
of the high-dimensional disorder-driven physics with magnetic field.}

\subsubsection*{Interplay of rare-region and perturbative effects.} In our view, it currently remains an open
question whether the non-Anderson transitions discussed here are true phase transitions or sharp crossovers.
This question requires the development of a field-theoretical description of disordered systems that accounts for
the interplay of perturbative and rare-region effects.

As discussed in Sec.~\ref{Sec:rare}, rare-region effects are qualitatively distinct in systems that allow
for localisation (semiconductors, Dirac semimetals, non-chiral chains of atoms, etc.)
and systems where localisation is forbidden
by symmetry and topology (chiral chains, single-node Weyl semimetals).


\subsubsection*{Models of infinite-dimensional systems ($d\rightarrow\infty$).} Because the unconventional
disorder-driven physics discussed in this review emerges in sufficiently high dimensions,
it is natural to expect manifestations of this physics in random regular graphs and
tree-like structures, which have been used to study the properties of the Anderson localisation 
transition in large and infinite
dimensions [see, e.g., Refs.~\cite{ACAT,MirlinFyodorov,MirlinFyodorov2,Dobrosavljevic,Tikhonov:RRG}].

\subsubsection*{Novel candidate materials.} Systems that exhibit non-Anderson transitions
and/or high-dimensional disorder-driven physics are probably not limited to those discussed
in this review and shown in Fig.~\ref{PeriodicTable}.

We believe that the possibility of observing these effects in other types of systems deserves further investigation.
Apparent candidates include, but are not limited to, excitations in interacting systems near quantum
phase transitions (e.g. the superconductor-insulator transition), photonic systems and phonons in disordered materials.

\section*{DISCLOSURE STATEMENT}
The authors are not aware of any affiliations, memberships, funding, or financial holdings that
might be perceived as affecting the objectivity of this review.

\section*{ACKNOWLEDGMENTS}
We are grateful to M.~G\"arttner, A.M.~Ostrovsky, A.M.~Rey, Ya.I.~Rodionov, B.~Sbierski, B.~Skinner
and especially V.~Gurarie
for collaboration on the subjects of this review
and to R.~Nandkishore, B.~Normand, T.~Ohtsuki, A.~Prem, B.~Sbierski and M.~Sch\"utt for comments on the manuscript.
We are also deeply indebted to B. Normand for the proofreading of the manuscript and numerous suggestions
and to B.~Skinner for insightful discussions of the review.
Our work was supported financially by NSF grants
DMR-1001240, DMR-1205303, PHY-1211914, PHY-1125844 {and PHY-1125915}.
LR also acknowledges a Simons Investigator award from the Simons Foundation
{and the hospitality of the Kavli Institute of Theoretical Physics where a part of this work was completed.}

\vspace{1cm}

\end{document}